\documentclass[aps,prl,twocolumn,showpacs,superscriptaddress]{revtex4}
\usepackage{epsfig}
\usepackage{graphicx}
\usepackage{amsfonts}
\usepackage[figuresright]{rotating}
\usepackage{amssymb}
\usepackage{amsmath}
\usepackage{psfrag}

\def\be{\begin{equation}}
\def\ee{\end{equation}}
\def\bea{\begin{eqnarray}}
\def\eea{\end{eqnarray}}

\def\nn{\nonumber}
\def\pp{\parallel}

\newcommand{\ket}[1]{| #1 \rangle}

\begin{document}
\title{High-Dimensional Topological Insulators with
Quaternionic Analytic Landau Levels}
\author{Yi Li}
\affiliation{Department of Physics, University of California, San Diego,
La Jolla, CA 92093}
\author{Congjun Wu}
\affiliation{Department of Physics, University of California, San Diego,
La Jolla, CA 92093}
%\affiliation{the Key Laboratory of Artificial Micro- and Nano-structures at the Ministry of
%Education, the School of Physics and Technology at Wuhan University, Wuhan 430072, China}

\begin{abstract}
We study the 3D topological insulators in the continuum by coupling
spin-$\frac{1}{2}$ fermions to the Aharonov-Casher SU(2) gauge field.
They exhibit flat Landau levels in which orbital angular momentum
and spin are coupled with a fixed helicity.
The 3D lowest Landau level wavefunctions exhibit the quaternionic analyticity
as a generalization of the complex analyticity of the 2D case.
Each Landau level contributes one branch of gapless helical Dirac
modes to the surface spectra, whose topological properties belong
to the $\mathbb{Z}_{2}$-class.
The flat Landau levels can be generalized to an arbitrary dimension.
Interaction effects and experimental realizations are
also studied.
\end{abstract}
\maketitle

The 2D quantum Hall (QH) systems \cite{Klitzing1980,tsui1982} are among the
earliest examples of quantum states characterized by topology
\cite{thouless1982,haldane1988} rather than symmetry in condensed
matter physics.
Their magnetic band structures possess topological Chern numbers defined
in time-reversal (TR) symmetry breaking systems \cite{thouless1982,avron1983,niu1985,kohmoto1985,avron1990}.
The consequential quantized charge transport originates from chiral edge
modes \cite{laughlin1981,halperin1982}, a result from the chirality of
Landau level wavefunctions.
Current studies of TR invariant topological insulators (TIs) have made great
success in both 2D and 3D.
They are described by a $\mathbb{Z}_{2}$-invariant which is topologically
stable with respect to TR invariant perturbations
\cite{bernevig2006a,kane2005,kane2005a,fu2007,fu2007a,moore2007,bernevig2006,
qi2008,roy2009,roy2010,hasan2010,qi2011}.
On open boundaries, they exhibit odd numbers of gapless helical edge modes in
2D systems and surface Dirac modes in 3D systems.
TIs have been experimentally observed through transport experiments
\cite{konig2007,qu2010,xiong2012} and spectroscopic measurements
\cite{roushan2009,hsieh2008,hsieh2009,xia2009,chen2009,alpichshev2010,
zhang2009a}.

The current research of 3D TIs has been focusing on the Bloch-wave band
structures.
Nevertheless, Landau levels (LLs) possess the advantages of the elegant analytic properties
and flat spectra, both of which have played essential roles in the study
of 2D integer and fractional QH effects \cite{laughlin1983,Haldane1983,halperin1984,girvin1984,arovas1984,
haldane1985,girvin1987,jain1989,jain2007, zhang1989,moore1991,wen1992,sondhi1993, murthy2003,DasSarma2005,DasSarma2008}.
As pioneered by Zhang and Hu \cite{zhang2001}, LLs and QH effects have
been generalized to various high dimensional manifolds
\cite{zhang2001,bernevig2002,elvang2003,fabinger2002,bernevig2003,hasebe2010}.
However, to our knowledge, TR invariant isotropic LLs have not been studied
in 3D flat space before.
It would be interesting to develop the LL counterpart of 3D TIs in the
continuum independent of the band inversion mechanism.
The analytic properties of 3D LL wavefunctions and the flatness of their
spectra provide an opportunity for further investigation on non-trivial
interaction effects in 3D topological states.

In this Letter, we construct 3D isotropic flat LLs in which spin-$\frac{1}{2}$
fermions are coupled to an $SU(2)$ Aharonov-Casher potential.
When odd number LLs are fully filled,
the system is a 3D $\mathbb{Z}_2$ TI with TR symmetry.
Each LL state has the same helicity structure, {\it i.e.},
the relative orientation between orbital angular momentum and spin.
Just like that the 2D lowest LL (LLL) wavefunctions in the symmetric gauge
are complex analytic functions, the 3D LLL ones are mapped
into quaternionic analytic functions.
Different from the 2D case, there is no magnetic translational
symmetry for the 3D LL Hamiltonian due to the non-Abelian nature
of the gauge field.
Nevertheless, magnetic translations can be applied for the
Gaussian pocket-like localized eigenstates in the LLL.
The edge spectra exhibit gapless Dirac modes.
Their stability against TR invariant perturbations indicates the
$\mathbb{Z}_{2}$ nature.
This scheme can be easily generalize to $N$ dimensions.
Interaction effects and
the Laughlin-like wavefunctions for the 4D case are  constructed.
Realizations of the 3D LL system are discussed.

We begin with the 3D LL Hamiltonian for a spin-$\frac{1}{2}$
non-relativistic particle as
\bea
H^{3D,LL}=\frac{1}{2m}
\sum_a
\big \{
 -i\hbar \nabla^a - \frac{q}{c} A^a (\vec{r})
\big\}^{2}
+V(r),
\label{eq:ham0}
\eea
where $A^a_{\alpha \beta }=\frac{1}{2}G \epsilon_{abc} \sigma_{\alpha \beta}^b
r^c$ is a 3D isotropic $SU(2)$ gauge with Latin indices run over $x,y,z$ and
Greek indices denote spin components $\uparrow, \downarrow$; $G$ is a coupling
constant and $\sigma$'s are Pauli matrices; $V(r)=-\frac{1}{2}m\omega _{0}^{2}r^{2}$
is a harmonic potential with $\omega _{0}=|qG|/(2mc)$ to maintain the flatness of LLs.
$\vec A$ can be viewed as an Aharonov-Casher
potential associated with a radial electric field
linearly increasing with $r$ as $\vec{E}(r)\times \vec{\sigma}$.
$H^{3D,LL}$ preserves the TR symmetry in contrast to the 2D QH
with TR symmetry broken.
It also gives a 3D non-Abelian generalization of the 2D quantum spin Hall
Hamiltonian based on Landau levels studied in Ref. \cite{bernevig2006a}.
More explicitly, $H^{3D,LL}$ can be further expanded as a harmonic
oscillator with a constant spin-orbit (SO) coupling as
\bea
H^{3D,LL}_{\mp }=\frac{p^{2}}{2m}+ \frac{1}{2}m\omega _{0}^{2}r^{2}
\mp \omega _{0}\vec{\sigma}\cdot \vec{L},
\label{eq:ham1}
\eea
where $\mp $ apply to the cases of $qG>0~(<0)$, respectively.
The spectra of Eq. \ref{eq:ham1} were  studied
in the context of the supersymmetric quantum mechanics \cite{bagchi2001}.
However, its connection with Landau levels was not noticed.
Eq. \ref{eq:ham0} has also been proposed to describe the
electrodynamic properties of superconductors
\cite{hirsch2008, hirsch2008a, hirsch2013}.

\begin{figure}[tbp]
\centering\epsfig{file=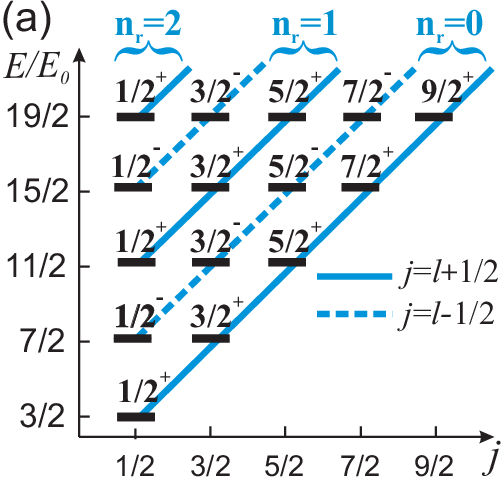,clip=1,width=0.48\linewidth,angle=0}
\hspace{2mm}
\centering\epsfig{file=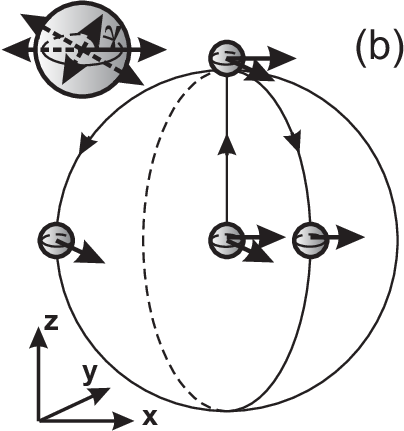,clip=1,width=0.42\linewidth,
angle=0}
\caption{a) The eigenstates of the 3D harmonic oscillator labeled by total
angular momentum $j_\pm=l\pm \frac{1}{2}$. Following the solid diagonal
(dashed) lines, these states are reorganized into the 3D
LL sates with the positive (negative) helicity.
b)
The magnetic translation for the LLL state ($l=0$) localized
at the origin in the case of $qG>0$, whose spin is set along
an arbitrary direction in the $xy$-plane.
The displacement vector $\vec \delta$ lies in the plane perpendicular
to spin orientation. The resultant state remains in the LLL as a
localized Gaussian pocket.
}
\label{fig:spec}
\end{figure}

The spectra and eigenstates of Eq. \ref{eq:ham0} are explained as follows.
We introduce the helicity number for the eigenstate of $\vec L \cdot \vec
\sigma$, defined as the sign of its eigenvalue
of the total angular momentum $\vec J=\vec L +\vec S$,
which equals $\pm 1$ for the sectors of
$j_\pm=l\pm \frac{1}{2}$, respectively.
At $qG>0$, the eigenstates are denoted as $\psi_{n_r; j_\pm, j_z; l}(\vec
r)=R_{n_r,l}(r) \mathcal{Y}_{j_\pm, j_z;l}(\hat \Omega)$, where the radial function
is $R_{n_{r}l}(r)= r^{l}e^{-\frac{r^{2}}{4l_G^{2}}}F(-n_{r},l+\frac{3}{2},
\frac{r^{2}}{2l_G^2})$; $F$ is the confluent hypergeometric function and
$l_G=\sqrt{\frac{\hbar c} {qG}}$ is the analogy of the
magnetic length;
$\mathcal{Y}_{j_\pm, j_z;l}(\hat \Omega)$'s are the spin-orbit
coupled spheric harmonic with $j_\pm=l\pm \frac{1}{2}$, respectively.
Flat spectra appear with infinite degeneracy in the
sector of $j_+$, where the energy dispersion
$E_{n_r,l}^+=(2n_r+\frac{3}{2})\hbar \omega_0$ is independent of $l$, and
thus $n_r$ serves as the LL index.
For the sector of $j_-$, the energy disperses with $l$
as $E_{n_r,l}^-=[2(n_r+l)+\frac{5}{2}]\hbar \omega_0$.
Similar results apply to the case of $qG<0$, where the
infinite degeneracy occurs in the sector of $j_-$.
These LL wavefunctions are the same as those of the 3D
harmonic oscillator but with different organizations.
As illustrated in Fig. \ref{fig:spec} (a),  these
eigenstates along each diagonal line with the positive
(negative) helicity fall into the flat LL states
for the case of $qG>0~(<0)$, respectively.
The ladder algebra generating the whole 3D LL states is explained in the Supplemental Material \cite{suppl}.

%-------------------------------------------------------------------
%------------------------------------------------------------------

Compared to the 2D case, a marked difference is that the 3D LL
Hamiltonian has no magnetic translational symmetry.
The non-Abelian field strength grows quadratically with $r$
as $F_{ij}(\vec{r})=\partial _{i}A_{j}-\partial _{j}A_{i}-\frac{iq}{\hbar c}%
[A_{i},A_{j}]=g\epsilon _{ijk}\big\{\sigma ^{k}+\frac{1}{4l_{G}^{2}}r^{k}(\vec{%
\sigma}\cdot \vec{r})\big\}$.
Nevertheless, magnetic translations still apply to the highest
weight states of the total angular momentum
$\vec J=\vec L +\vec S$ in the LLL at $qG>0$.
For simplicity, we drop the normalization factors of wavefunctions below.
For the positive helicity states with $j_z=j_+$,
$\vec L$ and $\vec S$ are parallel to each other.
Their wavefunctions are denoted by $\psi _{\hat{z}, l}^{hw}(\vec{r})
= (x+iy)^{l}e^{-\frac{r^2}{4l_G^2}}\otimes \alpha _{\hat{\Omega}= \hat{z}}$,
where $\alpha _{\hat{\Omega}}$ is the spin eigenstate of $\hat{\Omega}
\cdot \vec{\sigma}$ with eigenvalue $1$.
For these states, the magnetic translation is defined as usual
$T_{\hat{z}}(\vec{\delta})=\exp [-\vec \delta \cdot \vec{\nabla}
+\frac{i}{4l_{G}^{2}} \vec r_{xy} \cdot
(\hat z \times \vec \delta)]$,
where $\vec{\delta}$ is the displacement vector
in the $xy$-plane and $\vec{r}_{xy}$ is the projection
of $\vec r$ in the $xy$-plane.
The resultant state,
$T_{\hat{z}}(\vec{\delta})\psi _{\hat{z},l}^{hw}(\vec{r}
)=e^{i \frac{\vec r_{xy} \cdot (\hat z \times \delta)}{4l_{G}^{2}}}\psi _{
\hat{z},l}^{hw}(\vec{r}-\vec{\delta})$, remains in the LLL.
Generally speaking, the highest weight states can be
defined in a plane spanned by two orthogonal unit vectors
$\hat{e}_{1,2}$ as
$\psi _{\hat{e}_{3}, l}^{hw}(\vec{r})= \lbrack (\hat{e}_{1}+ i
\hat{e}_{2})\cdot \vec{r}]^{l}e^{-\frac{r^2}{4l_{G}^{2}}}
\otimes \alpha _{\hat{e}_{3}}$ with
$\hat{e}_{3}=\hat{e}_{1}\times \hat{e}_{2}$.
The magnetic translation for such states is defined as
$T_{\hat{e}_{3}}(\vec{\delta})=\exp [-\vec \delta \cdot \vec{\nabla}+
\frac{i}{4l_{G}^{2}}
~\vec r_{12} \cdot (\hat e_3 \times \vec{\delta})]$,
where $\vec{\delta}$ lies in the $\hat{e}_{1,2}$-plane and $\vec{r}_{12}=
\vec{r}-\hat{e}_{3}(\vec{r}\cdot \hat{e}_{3})$.
As an example, let us translate the LLL state localized at the origin
as illustrated in Fig. \ref{fig:spec} (b).
We set the spin direction of $\psi^{LLL}_{\hat e_3,l=0}$ in the $xy$%
-plane parameterized by $\hat e_3(\gamma)=\hat x\cos \gamma +\hat y\sin
\gamma$, \textit{i.e.}, $\alpha_{\hat e_3}(\gamma)= \frac{1}{\sqrt 2}
(|\uparrow \rangle +e^{i\gamma} |\downarrow\rangle)$, and translate it along
$\hat e_1=\hat z$ at the distance $R$.
The resultant states  read as
\bea
\psi_{\gamma,R} (\rho,\phi,z) =e^{i\frac{g}{2} R\rho \sin (\phi-\gamma)} e^{-|\vec r- R
\hat z|^2/4l_G^2}\otimes \alpha_{\hat e_3}(\gamma),
\label{eq:off-center}
\eea
where $\rho=\sqrt{%
x^2+y^2}$ and $\phi$ is the azimuthal angular of $\vec r$
in the $xy$-plane.
Such a state remains in the LLL as an off-centered Gaussian wave packet.

The highest weight states and their descendent states from
magnetic translations defined above have a clear classic
picture.
The classic equations of motion are derived as
\bea
\dot{\vec{r}}&=&\frac{1}{m}\vec{
p}+2\omega _{0}(\vec{r}\times \frac{1}{\hbar }\vec{S}), \ \ \,
\dot{\vec{p}} =2\omega _{0}\vec{p}\times \frac{1}{\hbar }
\vec{S}-m\omega _{0}^{2}\vec{r},
\nn \\
\dot{\vec{S}}&=& \frac{2\omega _{0}}{\hbar} \vec{S}\times \vec{L},
\label{eq:classic}
\eea
where $\vec p$ is the canonical momentum,
$\vec{L}=\vec{r}\times \vec{p}$ is the canonical orbital
angular momentum,
and
$\vec{S}$ here is the expectation value of $\frac{\hbar }{2} \vec \sigma$.
The first two describe the motion in a non-inertial frame
subject to the angular velocity $\frac{2\omega _{0}}{\hbar }\vec{S}$, and
the third equation is the Larmor precession.
$\vec L \cdot \vec S$ is a constant of motion of Eq. \ref{eq:classic}.
In the case of $\vec S \pp \vec L$, it is easy to prove that
both $\vec S$ and $\vec L$ are conserved.
Then the cyclotron motions become coplanar within the equatorial plane
perpendicular to $\vec S$.
Centers of the circular orbitals can be located at any points
in the plane.

The above off-centered LLL states break all the rotational symmetries.
Nevertheless, we can recover the rotational symmetry around the axis
determined by the origin and the packet center.
Let us perform the Fourier
transform of $\psi_{\gamma,R}(\rho,\phi,z)$ in Eq. \ref{eq:off-center}
with respect to the
azimuthal angle $\gamma$ of spin polarization.
The resultant state,
$\psi _{j_{z}=m+\frac{1}{2},R}(\rho ,\phi ,z)=\int_{0}^{2\pi }\frac{d\gamma
}{2\pi }e^{im\gamma }\psi _{\gamma ,R}$, is a $j_z$-eigenstate as
\begin{eqnarray}
e^{\frac{-|\vec{r}-R\hat{z}|^{2}}{4l_{G}^{2}}}
e^{im\phi }\Big\{J_{m}(x)|\uparrow \rangle
+J_{m+1}(x)e^{i\phi }|\downarrow \rangle \Big\},
\label{eq:fourier}
\end{eqnarray}
with $x=R\rho /(2l_{G}^{2})$.
At large distance of $R$, the spatial extension of $\psi_{j_{z}=m+\frac{1}{2},R}$
in the $xy$-plane is at the order of $ml_G^2/R$, which is suppressed
at large values of $R$ and scales linear with $m$.
In particular, the narrowest states $\psi_{\pm \frac{1}{2},R}$ exhibit an
ellipsoid shape with an aspect ratio decaying as $l_G/R$
when $R$ goes large.
%For those states with $|m|<R/l_{G}$, they localize within the
%distance of $l_{G}$ from the center $R\hat{z}$.
%As a result, the real space local density of states of LLL grow linearly
%with $R$.
%In fact, this local density of state can be calculated exactly as
%$\rho(r)=\frac{1}{\pi l_G^3}
%\Big\{ \frac{1}{\sqrt \pi} e^{-\frac{r^2}{l_G^2}}
%+(\frac{r}{l_G}+\frac{l_G}{2r})\mbox{erf}
%(\frac{r}{l_G})\Big\}$, which approaches
%$\frac{r}{\pi l_G^4}$ as $r\rightarrow +\infty$.

In analogy to the fact that the 2D LLL states are complex analytic
functions due to chirality, we have found an impressive result
that the helicity in 3D LL systems leads to the quaternionic analyticity.
Quaternion is the first discovered non-commutative division algebra,
which has three anti-commuting imaginary units
$i$, $j$ and $k$, satisfying
$i^2=j^2=k^2=-1$ and $ij=k$. %, $jk=i$, and $ki=j$.
It has been applied in quantum systems \cite{Balatsky1992, adler1995}
and SO coupled Bose-Einstein condensations \cite{li2012b}.
Just like two real numbers forming a complex number,
a two-component complex spinor $\psi=(\psi_\uparrow, \psi_\downarrow)^T$ can
be viewed as a quaternion defined as $f=\psi_\uparrow + j \psi_\downarrow$.
In the quaternion representation, the TR transformation
$i\sigma_2 \psi^*$ becomes $T f=-fj$ satisfying $T^2=-1$;
multiplying a U(1) phase factor $e^{i\phi} \psi$ corresponds to $f e^{i\phi}$;
the SU(2) operations $e^{-i\frac{\sigma_x}{2}\phi}\psi$,
$e^{-i\frac{\sigma_y}{2}\phi}\psi$, and $e^{-i\frac{\sigma_z}{2}\phi}\psi$
map to $e^{\frac{k}{2} \phi}f$, $e^{\frac{j}{2}\phi}f$,
and $e^{-\frac{i}{2}\phi}f$, respectively.
The quaternion version of
$\psi^{LLL}_{j=j_+, j_z=m+\frac{1}{2}}$ is $f_{j_+,j_z}^{LLL}(x,y,z)=
\Psi_{\uparrow,j_+,j_z}+ j\Psi_{\downarrow,j_+,j_z}$,
where
$\Psi_{\uparrow,j_+,j_z}=\langle j_+, j_z | l,m;\frac{1}{2},\frac{1}{2}\rangle
r^l Y_{l,m}$, $ \Psi_{\downarrow,j_+,j_z}= \langle j_+, j_z | l,m+1;\frac{1}{2},
-\frac{1}{2}\rangle r^l Y_{l,m+1}$.
Please note that the Gaussian factor does not appear in $f_{j_+,j_z}^{LLL}$
which is a quaternionic polynomial.

As a generalization of the Cauchy-Riemann condition, a quaternionic
analytic function $f(x,y,z,u)$ satisfies the Fueter condition
\cite{sudbery1979} as
\bea
\frac{\partial f}{\partial x}
+i\frac{\partial f}{\partial y}
+j\frac{\partial f}{\partial z}
+k\frac{\partial f}{\partial u}=0,
\label{eq:quaternion}
\eea
where $x$, $y$, $z$ and $u$ are coordinates in the 4D space.
In Eq. \ref{eq:quaternion}, imaginary units are multiplied
from the left, thus it is the left-analyticity condition
which works in our convention.
Below, we prove the LLL function $f_{j_+,j_z}^{LLL}(x,y,z)$ satisfying
Eq. \ref{eq:quaternion}.
Since $f_{j_+,j_z}^{LLL}$ is defined in 3D space, it is a constant over
$u$, and thus only the first three terms in Eq. \ref{eq:quaternion}
apply to it.
Obviously the highest weight states
with spin along the $z$-axis,
$f^{LLL}_{j_+=j_z=l+\frac{1}{2}}=(x+ i y)^l$, satisfy Eq. \ref{eq:quaternion}
which is reduced to complex analyticity.
By applying an arbitrary SU(2) rotation $g$
characterized by the Eulerian
angles $(\alpha, \beta, \gamma)$,
$f^{LLL}_{j_+=j_z}$ transforms to
\bea
f^{\prime,~ LLL}(x,y,z)
=e^{-i\frac{\alpha}{2}} e^{j\frac{\beta}{2}}
e^{-i\frac{\gamma}{2}} f^{LLL}_{j_+=j_z}( x^\prime, y^\prime, z^\prime),
\label{eq:rotation}
\eea
where  $(x^\prime,y^\prime$, $z^\prime)$ are the coordinates
by applying the inverse of $g$ on $(x,y,z)$.
We check that
$(\frac{\partial}{\partial x}+ i\frac{\partial}{\partial y}
+j\frac{\partial}{\partial z}) f^{\prime LLL}(x,y,z)
=e^{i\frac{\alpha}{2}} e^{-j\frac{\beta}{2}} e^{i\frac{\gamma}{2}}
\Big\{\frac{\partial}{\partial x'} +i\frac{\partial}{\partial y'}
+j\frac{\partial}{\partial z'} \Big\}
f^{LLL}_{j_+,j_z}(x^\prime, y^\prime, z^\prime)=0$.
Essentially, we have proved that Fueter condition is rotationally
invariant.
Since all the highest weight states are connected through SU(2) rotations,
and they form an over-complete basis for
the angular momentum representations, we conclude that all the 3D LLL
states with the positive helicity are quaternionic analytic.

Next we prove that the set of quaternionic LLL states $f^{LLL}_
{j_+=l+\frac{1}{2},j_z}$%(j_+\ge \frac{1}{2}, j_z\ge j_+\ge -j_z)
form the complete basis for quaternionic valued analytic
polynomials in 3D.
Any linear superposition of the LLL states with $j_+$
can be represented as
$f_l =\sum_{j_z=-j_+}^{j_+} f^{LLL}_{j_+,j_z}  ~c_{j_z}$,
where $c_{j_z}$ is a complex coefficient.
Because of the TR relation $f^{LLL}_{j_+,-j_z}=-f^{LLL}_{j_+,-j_z}j$,
$f_l$ can be expressed in terms of $l+1$
linearly independent basis as
\bea
f_l(x,y,z) =\sum_{m=0}^{l} f^{LLL}_{j_+=l+\frac{1}{2},j_z=m+\frac{1}{2}} q_m,
\label{eq:complete}
\eea
where $q_m= c_{m+\frac{1}{2}}-j c_{-m-\frac{1}{2}}$ is a quaternion constant.
On the other hand,  it can be calculated that
the rank of the linearly independent
$l$-th order quaternionic polynomials satisfying Eq. \ref{eq:quaternion}
is just $C^2_{l+2}-C^2_{l+1}=l+1$, thus $f^{LLL}_{j_+,j_z}$'s with
$j_z\ge \frac{1}{2}$ are complete.

The topological nature of the 3D LL problem exhibits clearly in the gapless
surface states.
A numeric calculation of the gapless surface spectra is presented in the
Supplemental Material \cite{suppl}.
At $qG>0$, inside the bulk, LL spectra are flat with respect to
$j_+=l+\frac{1}{2}$.
As $l$ goes large, the classical orbital radius $r_c$ approaches the
open boundary with the radius $R_0$.
For example, for a LLL state, $r_{c}=\sqrt{2l}l_{G}$.
States with $l> l_{c}\approx \frac{1}{2}(R_{0}/l_{G})^{2}$ become surface states.
Their spectra become
$E(l)\approx l(l+1)\frac{\hbar ^{2}}{2mR_{0}^{2}}-l\hbar \omega _{0}$.
When the chemical potential $\mu$ lies inside the gap, it cuts the surface
states with the Fermi angular momentum denoted by $l_{f}$.
These surface states satisfy $\vec \sigma\cdot \vec L=l\hbar$, thus the
their spectra can be linearized around $l_{f}$ as
$H_{bd}=(v_{f}/R_{0})\vec{\sigma} \cdot \vec{L}-\mu $.
This is the Dirac equation defined on a sphere with the radius $R_{0}$.
It can be expanded around $\vec{r}=R_{0}\hat{e}_{r}$ as $H_{bd}=\hbar v_{f}(%
\vec{k}\times \vec{\sigma})\cdot \hat{e}_{r}-\mu $. Similar reasoning
applies to other Landau levels which also give rise to Dirac spectra.
Because of the lack of Bloch wave band structure, it remains a
challenging problem to directly calculate the bulk topological index.
Nevertheless, the $\mathbb{Z}_2$ structure manifests through the
surface Dirac spectra.
Since each fully occupied LL contributes one helical Dirac Fermi surface,
the bulk is $\mathbb{Z}_2$-nontrivial (trivial) if odd (even) number of LLs
are occupied.
In the $\mathbb{Z}_2$-nontrivial case, the gapless helical surface states
are protected by TR symmetry and are robust under TR invariant perturbations.

In Eq. 2, the harmonic frequency $\omega_T$ is set to be equal
with the SO frequency $\omega_0$ to maintain the flatness of LL spectra.
However, the $\mathbb{Z}_2$ topology of the 3D LLs does not rely on this.
Define $\Delta \omega=\omega_T -\omega_0$, and we set $\Delta\omega\ge0$ to
maintain the spectra bounded from below.
$\Delta \omega>0$ corresponds to imposing an external potential
$\Delta V(r)=\frac{1}{2}m (\omega_T^2-\omega_0^2) r^2$ to the bulk
Hamiltonian of Eq. 2.
If $\Delta \omega\ll \omega_0$, $\Delta V(r)$ is soft.
It results in energy dispersions of 3D LLs but does not affect their topology.
For simplicity, let us check the case of $qG>0$.
The $\vec \sigma\cdot \vec L$ term commutes with the overall harmonic
potential, thus the LL wavefunctions remain the same as those of
Eq. 2 by replacing $\omega_0$ with $\omega_T$.
Their dispersions become
$E^+_{n_r,j_+}=(2n_r + 1)\hbar\omega_T +\frac{1}{2}\hbar\omega_0
+j_+\hbar \Delta\omega$ which are very slow.
In other words, $\Delta V(r)$ imposes a finite sample size with the radius of
$R^2<\hbar /(m\Delta \omega)=2l_G \frac{\omega_0} {\Delta\omega}$ even
without an explicit boundary.
Inside this region, $\Delta V$ is smaller than the LL gap, and the LL
states are bulk states.
Their energies are within the LL gap and the angular momentum
numbers $j_+<\frac{2\omega_T}{\Delta\omega}$.
LL states outside this region can be viewed as surface states with positive
helicity.
For a given Fermi energy, it also cuts a helical Fermi surface with the
same form of effective surface Hamiltonian.

The above scheme can be easily generalized to arbitrary dimensions \cite{suppl} by
combining the $N$-D harmonic oscillator potential and SO coupling.
For example, in 4D, we have
$H^{4D, LL}=\frac{p_{4D}^2}{2m}+\frac{1}{2}m \omega_0^2 r_{4D}^2-\omega_0
\sum_{1\le a<b \le 4} \Gamma^{ab} L_{ab}$, where $L_{ab}=r_a p_b - r_b p_a$
and the 4D spin operators are defined as
$\Gamma^{ij}=-\frac{i}{2}[\sigma^{i}, \sigma^{j}]$,
$\Gamma^{i4}=\pm \sigma^{i}$ with $1\le i<j\le3$.
The $\pm$ signs of $\Gamma^{i4}$ correspond to two complex
conjugate irreducible
fundamental spinor representations of $SO(4)$, and the $+$
sign will be taken below.
The spectra of the positive helicity states are flat as
$E_{+, n_r}=(2n_{r}+2) \hbar \omega$.
Following a similar method in 3D, we prove that the quaternionic
version of the 4D LLL wavefunctions satisfy the full equation of
Eq. \ref{eq:quaternion}.
They form the complete basis for quaternionic left-analytic polynomials
in 4D.
For each $l$-th order, the rank can be calculated as
$C^3_{l+3}-C^3_{l+2}=\frac{1}{2}(l+1)(l+2)$.

We consider the interaction effects in the LLLs.
For simplicity, let us consider the 4D system and the short-range
interactions.
Fermions can develop spontaneous spin polarization to minimize the
interaction energy in the LLL flat band.
Without loss of generality, we assume that spin takes the eigenstate of
$\Gamma^{12}=\Gamma^{34}=\sigma^3$ with the eigenvalue 1.
The LLL wavefunctions satisfying this spin polarization can be expressed
as $\Psi^{LLL,4D}_{m,n}=(x+iy)^m (z+iu)^n e^{-\frac{r^2_{4D}}{4l_G^c2}}
\otimes \ket{\alpha}$ with $\ket{\alpha}=(1,0)^T$.
The 4D orbital angular momentum number for the orbital wavefunction
is $l=m+n$ with $m\ge 0$ and $n\ge 0$.
It is easy to check that  $\Psi^{LLL,4D}_{m,n}$ is the eigenstate of
$\sum_{ab} L_{ab}\Gamma^{ab}$ with the eigenvalue $(m+n)\hbar$.
If all the $\Psi^{LLL,4D}_{m,n}$'s are filled with $0 \le m < N_m$ and
$0 \le n < N_n $, we write down a Slater-determinant wavefunction as
\bea
\Psi(v_1, w_1; \cdots; v_N, w_N)= \det[v_i^{\alpha} w_i^{\beta}],
\label{eq:wf}
\eea
where the coordinates of the $i$-th particle form two pairs
of complex numbers abbreviated as $v_i=x_i+iy_i$ and $w_i=z_i+iu_i$;
$\alpha$, $\beta$ and $i$ satisfy $0 \le \alpha <N_m $,  $0\le
\beta < N_n $, and $1\le i \le N=N_m N_n$.
Such a state has a 4D uniform density as $\rho=\frac{1}{4\pi^4 l_G^4}$.
We can write down a Laughlin-like wavefunction
as the $k$-th power of Eq. \ref{eq:wf} whose
filling relative to $\rho$ should be $1/k^2$.
For the 3D case, we also consider the spin polarized
interacting wavefunctions.
However, it corresponds to that fermions concentrate to
the highest weight states in the equatorial plane
perpendicular to the spin polarization, and thus reduces
to the 2D Laughlin states.
In both 3D and 4D cases, fermion spin polarizations are spontaneous,
thus low energy spin waves should appear as low energy excitations.
Due to the SO coupled nature, spin fluctuations couple
to orbital motions, which leads to SO coupled excitations
and will be studied in a later
publication.

One possible experimental realization for the 3D LL system is the
strained semiconductors.
The strain tensor $\epsilon_{ab}=\frac{1}{2}(\partial_a u_b
+\partial_b u_a)$ generates SO coupling as $H_{SO}= \hbar\alpha
[(\epsilon_{xy}k_y -\epsilon_{xz} k_z) \sigma_x + (\epsilon_{zy}k_z
-\epsilon_{xy} k_x) \sigma_y +(\epsilon_{zx}k_x -\epsilon_{yz} k_y)
\sigma_z]$ where $\alpha=8\times 10^5$m/s for GaAs.
The 3D strain configuration with $\vec u=\frac{f}{2}(yz,zx,xy)$ combined
with a suitable scalar potential gives rise to Eq. \ref{eq:ham0}
with the correspondence $\omega_0=\frac{1}{2}\alpha f$.
A similar method was proposed in Ref. \cite{bernevig2006a} to realize
2D quantum spin Hall LLs.
A LL gap of 1mK corresponds to a strain gradient of the order of
1\% over 60 $\mu$m, which is accessible in experiments.
Another possible system is the ultra-cold atom system.
For example, recently evidence of fractionally filled 2D LLs with
bosons has been reported in rotating systems \cite{Gemelke2010}.

Furthermore,
synthetic SO coupling generated through atom-light interactions has
become a major research direction in ultra-cold atom system
\cite{lin2011,dalibard2011}.
The SO coupling term in the 3D LL Hamiltonian $\omega \vec \sigma
\cdot \vec L$ is equivalent to the spin-dependent Coriolis forces
from spin-dependent rotations, i.e., different spin eigenstates along
$\pm x$, $\pm y$ and $\pm z$ axes feel angular velocities
parallel to these axes, respectively.
An experimental proposal to realize such an SO coupling
has been designed and will be reported in a later publication
\cite{zhouprep}.

In conclusion, we have generalized the flat LLs to 3D and 4D flat spaces,
which are high dimensional topological insulators in the continuum
without Bloch-wave band structures.
The 3D and 4D LLL wavefunctions in the quaternionic version
form the complete bases of the quaternionic analytic
polynomials.
Each filled LL contributes one helical Dirac Fermi surface
on the open boundary.
The spin polarized Laughlin-like wavefunction is constructed
for the 4D case.
Interaction effects and topological excitations inside the LLLs in
high dimensions would be interesting for further investigation.
In particular, we expect that the quaternionic analyticity would
greatly facilitate this study.

This work grew out of collaborations with J. E. Hirsch, to whom
we are especially grateful.
Y. L. and C. W. thank S. C. Zhang and J. P. Hu for
helpful discussions.
Y. L. and C. W. are supported by NSF DMR-1105945 and
AFOSR. FA9550-11-0067 (YIP).

Note Added: Near the completion of this manuscript,
we learned that the 3D Landau level problem is also
studied by Zhang \cite{zhangprivate}.

%\bibliographystyle{prsty}
%\bibliography{TI}

%%%%%%%%%%%%%%%%%%%%%%%%%%%%%%%%%%%%%%%%%%%%%%%%%%%%%%%%%%%%%%%%%%%%
\newpage

\appendix

\begin{center}{\Large{Supplemental Material
for ``Topological insulators with quaternionic
analytic Landau levels''}}\end{center}

In this supplementary material, we present several points which are
not discussed in the maintext.
These include the ladder algebra to explain the degeneracy of the 3D
Landau level (LL) wavefunctions, the quaternionic version of the 3D
lowest Landau level (LLL) states with the negative helicity,
the numerical calculation on the 3D LL spectra with the open
boundary, and the generalization of LLs to an arbitrary dimension.

%-------------------------------------------------------------------
\begin{figure}[tbp]
\centering\epsfig{file=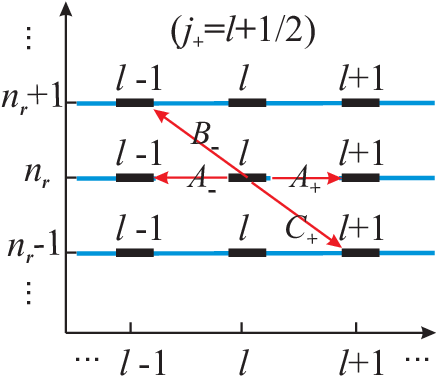,clip=1,width=0.8\linewidth, angle=0}
\caption{
The algebra structure of the 3D Landau levels in the positive helicity sector.
Operators $A_\pm(l)$ connect states with different $l$ in the same Landau
level, while $B_-(l)$ and $C_+(l)$ connect those between neighboring
Landau levels.
}
\label{fig:ladder}
\end{figure}
%------------------------------------------------------------------

\section{The ladder algebra for the spectra flatness}
The degeneracy of the 3D LL over different values of angular momentum
is not accidental, but protected by a ladder algebra constructed below.
For example, we take the case of $qG>0$ and consider the positive helicity
Landau level states of $H_+$.
The variable transformation for the radial eigenstates is applied as $\chi
_{n_{r},l}(r)=rR_{n_{r},l}(r)$,
and the corresponding radial Hamiltonians become
\bea
H_{l}=\hbar \omega_0 \Big\{-\frac{d^{2}}{dr^{\ast 2}}+\frac{l(l+1)}{r^{\ast 2}}
-l+\frac{1}{4} r^{\ast 2} \Big\},
\eea
where the dimensionless radius is  $r^{\ast }=\frac{r}{l_{G}}$.
The ladder operators are defined as
\bea
A_{+}(l)&=&\frac{d}{dr^{\ast }}-\frac{l+1}{r^{\ast }}-\frac{1}{2}r^{\ast }, \nn \\
A_{-}(l)&=&-\frac{d}{dr^{\ast }}-\frac{l}{r^{\ast }}-\frac{1}{2}r^{\ast }.
\eea
They satisfy the relations
\bea
H_{l\pm 1}A_{\pm }(l)=A_{\pm }(l)H_{l}.
\eea
Consequentially, $\chi _{n_{r},l\pm 1}= A_{\pm }(l)\chi _{n_{r},l}$ with
the same energy independent of $l$.
All the states in the same LL can be reached by
successively applying $A_{\pm }$ operators.

To connect  different LLs, other two ladder operators are
defined as
\bea
B_{-}(l)&=&-\frac{d}{dr^{\ast }}-\frac{l}{r^{\ast}}
+\frac{1}{2}r^{\ast }, \nn \\
C_{+}(l)&=&\frac{d}{dr^{\ast }}-\frac{l+1}{r^{\ast }}+\frac{1}{2}r^{\ast },
\eea
which satisfy
\bea
H_{l-1}B_{-}(l)&=&B_{-}(l)(H_{l}+2\hbar \omega_0),\nn \\
H_{l+1}C_{+}(l)&=&C_{+}(l)(H_{l}-2\hbar \omega_0),
\eea
respectively.
By applying $B_{-}(l)$ ($C_{+}(l)$) to $\chi _{n_{r},l}(r)$, we arrive at
\bea
\chi
_{n_{r}+1,l-1}&=&B_{-}(l)\chi _{n_{r},l}, \nn \\
\chi
_{n_{r}-1,l+1}&=&C_{+}(l)\chi _{n_{r},l},
\eea
where the energy shifts $\pm 2\hbar
\omega _{0}$, respectively, as illustrated in Fig. \ref{fig:ladder}.
Similar algebra can also be constructed for the case of $qG<0$.

%-------------------------------------------------------------------
\begin{figure}[tbp]
\centering
\epsfig{file=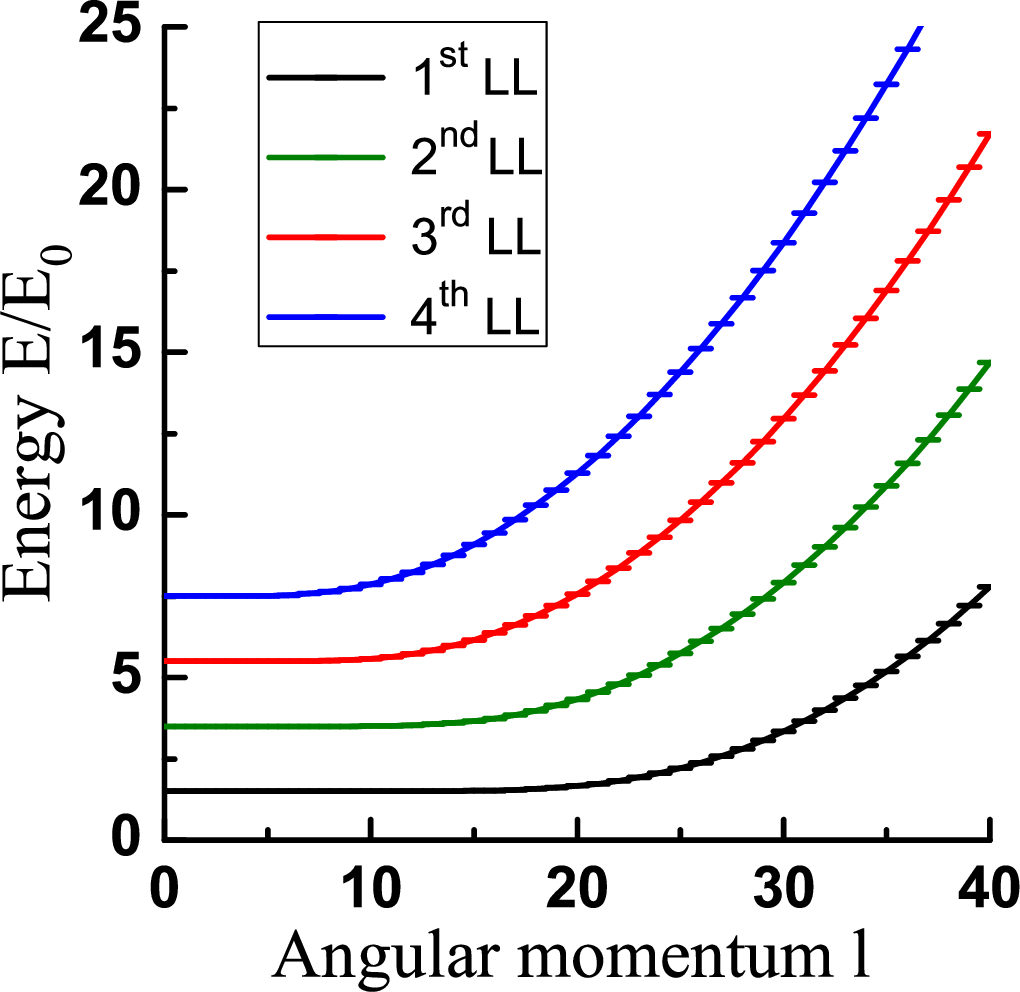,clip=2,width=0.7\linewidth, angle=0}
\caption{The energy dispersion of the first four Landau levels {\it v.s.}
$l=j-\frac{1}{2}$. Open boundary condition is used for a ball with the radius
$R_0/l_G=8$. The edge states correspond to those with large values of $l$ and
develop linear dispersions with $l$. The most probable radius of the LLL
state with $l$ is $r=l_G \protect\sqrt l$. }
\label{fig:edge}
\end{figure}
%------------------------------------------------------------------

\section{Numerical calculation for the gapless surface Dirac modes }
The surface states of the 3D LL Hamiltonian Eq. 2 in the main text
are gapless helical Dirac modes.
We have numerically calculated the spectra
with the open boundary condition for the positive helicity states
with $j_+=l+\frac{1}{2}$.
The results for the first four LLs are plotted in Fig. \ref{fig:edge}.
For the lowest LL (LLL) states, when the orbital angular momentum
$l$ exceeds a characteristic value $l_{c}\approx 30$, the spectra become
dispersive indicating the onset of surface states.
Although the surface spectra look very similar to those of the 2D quantum
Hall edges, a crucial difference is that each $l$
in Fig. \ref{fig:edge} does not represent a single chiral state
but a set of helical states of $2j_++1$ fold degeneracy
with $j_+=l+\frac{1}{2}$.

\section{Quaternionic wavefunction for the $j_-$ sector~~}
In the main text, we have showed that the 3D LLL states with the positive
helicity in the quaternion representation form a set of complete basis for
the quaternionic left-analytic polynomials.
For the case of the LLL with negative helicity,
their quaternionic version $g^{LLL}_{j_-,j_z}(x,y,z)$
are not analytic any more.
Nevertheless, since the wavefunctions of the negative helicity
sector can be related to the positive one via
$\mathcal{Y}_{j_-, j_z; l+1}(\hat \Omega)
=-\vec{\sigma} \cdot \hat{\Omega}
\mathcal{Y}_{j_+, j_z; l}(\hat \Omega)$,
where $\vec{\sigma} \cdot \hat{\Omega}$ has odd parity,
their quaternionic version is related to the analytic one through
$g^{LLL}_{j_-=(l+1)-\frac{1}{2},j_z}
=(-\hat{x}k-\hat{y}j+\hat{z}i)f^{LLL}_{j_+=l+\frac{1}{2},j_z}i$.
Here, the representation of quaternion imaginary units $\{ i,j,k\}$
as Pauli matrices $\{i \sigma_z, -i \sigma_y, -i \sigma_x \}$ are used,
as derived from the rotational properties of the
spin-orbit coupled spheric harmonics.

%------------------------------------------------
\section{Generalization to $N$-dimensions}
The study in 3D and 4D LL systems can be generalized to $N$-D
by replacing the vector and scalar potentials in Eq. 1 in
the main text with the $SO(N)$ gauge field
\bea
A^a(\vec r) = g r^b S^{ab}, \ \ \,
V(r)=-\frac{N-2}{2} m\omega_0 r^2,
\eea
respectively, where $S^{ab}$
are the $SO(N)$ spin operators constructed based on the Clifford algebra.
The rank-$k$ Clifford algebra contains $2k+1$ matrices with the
dimension $2^k\times 2^k$ which anti-commute with each other denoted as
$\Gamma^a ~(1\le a\le 2k+1)$.
Their commutators generate
\bea
\Gamma^{ab}=-\frac{i}{2} [\Gamma^a, \Gamma^b],
\eea
for $1\le a<b\le 2k+1$.
For odd dimensions $N=2k+1$, the $SO(N)$ spin operators in the
fundamental spinor representation  can be constructed by using
the rank-$k$ matrices as $S^{ab}=\frac{1}{2}\Gamma^{ab}$.
For even dimensions $N=2k+2$, we can select $2k+2$ ones
among the $2k+3$ $\Gamma$-matrices of rank-$(k+1)$ to form
$S^{ab}=\frac{1}{2}\Gamma^{ab}$, then
all of $S^{ab}$ commute with $\Gamma^{2k+3}$.
This $2^{k+1}$-D spinor representation of $S^{ab}$ is thus reducible
into the fundamental and anti-fundamental representations.
Both of them are $2^k$-D, which can be constructed
from the rank-$k$ $\Gamma$-matrices
as $S^{a,2k+2}=\pm\frac{1}{2}\Gamma^{a} (1\le a \le 2k+1)$ and
$S^{ab}=\frac{1}{2}\Gamma^{ab} (1\le a < b< 2k+1)$, respectively.

As for TR properties, $\Gamma^a$'s are TR even and odd at
even and odd values of $k$, respectively.
We conclude that at $N=2k+1$, the $N$-D version
of the LL Hamiltonian is TR invariant in the fundamental spinor
representation.
At $N=4k$, it is also TR invariant in both the fundamental
and anti-fundamental representations.
However $N=4k+2$, each one of the fundamental and anti-fundamental
representations is not TR invariant, but transforms into each other
under TR operation.

Similarly, the $N$-D LL Hamiltonian can be reorganized
as the harmonic oscillator with SO coupling.
For the case of $qG>0$, it becomes
\bea
H_{N,+}=\frac{p^{2}}{2m}+\frac{1}{2} m\omega_0^{2}r^{2}-\hbar\omega_0\Gamma _{ab}L_{ab},
\eea
where $L_{ab}=r_ap_b -r_b p_a$ with $1\le a < b\le N$.
The $l$-th order $N$-D spherical harmonic functions are eigenstates
of $L^2=L_{ab} L_{ab}$ with the eigenvalue of $\hbar^2 l (l+D-2)$.
The $N$-$D$ harmonic oscillator has the energy spectra of
$E_{n_{r},l}=(2n_{r}+l+N/2)\hbar \omega$.
When coupling to the fundamental spinors, the $l$-th spherical harmonics
split into the positive helicity ($j_+$) and negative helicity $(j_-)$
sectors, whose eigenvalues of the $\Gamma_{ab} L_{ab}$ are
$\hbar l$ and $-\hbar (l+N-2)$, respectively.
For the positive helicity sector, its spectra become
independent of $l$ as $E_{+}=(2n_{r}+N/2)\hbar \omega$,
with the radial wave functions are
\bea
R_{n_{r}l}(r)= r^{l}e^{-r^{2}/4l_{G}^{2}}F(-n_{r},l+N/2,r^{2}/2l_{G}^{2}).
\eea
The highest weight states in the LLL can be written as
\bea
\psi _{ab,\pm l}^{hw}(\vec{r})= \lbrack (\hat{e}_{a}\pm i\hat{e}_{b})\cdot \vec{r}
]^{l}e^{-r^{2}/4l_{G}^{2}}\otimes \alpha_{\pm, ab},
\eea
where $\alpha_{\pm, ab}$ is the eigenstate of $\Gamma_{ab}$ with eigenvalue $\pm 1$, respectively.
The magnetic translation in the $ab$-plane by the displacement
vector $\vec \delta$ takes the form
\bea
T_{ab}(\vec{\delta})=\exp \big [-\vec \delta \cdot \vec{\nabla}+
\frac{i}{2l_{G}^{2}} \Gamma_{ab} (r_a \delta_b -r_b\delta_a)\big ].
\eea
Similarly to the 3D case, staring from the LLL state localized around
the origin with $l=0$, we can perform the magnetic translation
and Fourier transformation with respect to the transverse spin polarization.
The resultant localized Gaussian pockets are LLL states of
the eigenstates of the $SO(N-1)$ symmetry with respect
to the translation direction $\vec\delta$.
Again each LL contributes to one channel of surface Dirac modes on $S^{N-1}$
described by $H_{bd}=(v_{f}/R_{0})\Gamma _{ab}L_{ab}-\mu$.

\end{document}